\documentclass[aps,preprint,showpacs,floatfix,twocolumn,10pt]{revtex4}
\usepackage{graphicx}
\begin{document}
\title{Phase transitions of a single polyelectrolyte in a poor
solvent with explicit counterions}
\author{Anoop Varghese}
\email{anoop@imsc.res.in}
\author{Satyavani Vemparala}
\email{vani@imsc.res.in}
\author{R. Rajesh}
\email{rrajesh@imsc.res.in}
\affiliation{The Institute of Mathematical Sciences, C.I.T. Campus,
Taramani, Chennai 600113, India}
\date{\today}
\begin{abstract}
Conformational properties of a single flexible polyelectrolyte chain in a
poor solvent  are 
studied using constant temperature molecular dynamics simulation. 
The effects of counterions are explicitly taken in to account. Structural 
properties of various phases and the transition between these phases are 
studied by tracking the values of asphericity, radius of gyration, 
fraction of condensed counterions, number of non-bonded neighbours and 
Coulomb interaction energies. From our simulations,
we find strong evidence for a first-order phase transition from extended to
collapsed phase consistent with earlier theoretical predictions. We also
identify a continuous phase transition associated with the condensation
of counterions and estimate the critical exponents associated with the
transition. Finally, we argue that previous suggestions of existence of
an independent intermediate phase between extended and collapsed phases is
only a finite size effect. 
\end{abstract} 
\maketitle

\section{Introduction}

Polyelectrolytes are polymers, which in a polar solvent, release 
counterions into the solution, making the polymer backbone charged. 
Examples of polyelectrolytes include sulfonated polystyrene, 
polymethacrylic acid, DNA, RNA and proteins. Though the system is
 overall charge neutral, 
long range Coulomb interactions make the behaviour of 
polyelectrolytes considerably different from that of neutral 
polymers~\cite{dob,yan}. The static and dynamic properties of 
polyelectrolytes depend on the nature of the solvent, linear charge 
density of the polymer back bone, valency of the counterions, 
temperature, salt concentration, hydrodynamic interactions and the 
bending rigidity~\cite{dob,Netz}.

A polyelectrolyte chain in a poor solvent undergoes a series of phase 
transitions as the linear charge density of the chain is varied keeping 
temperature fixed. The rich phase diagram resulting from a competition 
between the long range Coulomb interaction and the short range excluded 
volume interaction has been studied theoretically 
~\cite{kar1,kar2,rub,Lyulin,liao,man,muthu1,pincus}, numerically 
~\cite{rub,Lyulin,liao,wink,micka,hans,jaya} and 
experimentally~\cite{Geissler,Williams,Tinoco,Boue,Klenin,Colby,muthu2}. At 
very low charge densities, the polyelectrolyte chain is in a collapsed 
phase, with the counterions uniformly distributed throughout the 
solution \cite{jaya}. On increasing the charge density, the 
polyelectrolyte chain makes a transition into the pearl-necklace phase 
in which the chain has two (dumbbell) or more globules connected through 
a string of monomers \cite{kar1,kar2,rub,Lyulin}. This instability of the 
charged collapsed phase is similar to the Rayleigh instability of a 
charged droplet~\cite{rayleigh,deserno}. While the pearl-necklace phase has been 
observed in numerical simulations~\cite{hans,yethi,jaya}, the 
experimental status is unclear~\cite{Geissler,Boue,Colby}. Further increase in 
charge density decreases the number and size of the globules and the 
polyelectrolyte chain becomes extended. When the linear charge density 
exceeds a threshold value, the electrostatic energy dominates over the 
thermal fluctuations and the counterions condense on the polyelectrolyte 
chain~\cite{man}. The condensed counterions form dipoles with the 
polymer monomers and the attraction among the dipoles leads to the 
collapse of the polyelectrolyte chain~\cite{wink}. Recent simulations 
also show that, in extreme poor solvent conditions, the polyelectrolyte 
chain can make a direct transition from the initial collapsed phase to 
the final condensed collapsed phase with out encountering some or even 
all the intermediate phases~\cite{jaya}.

Certain features of the phase diagram remain poorly understood. Recent 
simulations argue for the existence of an intermediate phase between the 
extended and the condensed collapsed phases \cite{jaya,hans}. Referred 
to as the sausage phase, this phase is defined as a collapsed phase in 
which the shape of the collapsed polymer becomes aspherical, having a 
non-zero mean asphericity. It is not clear whether this phase is just a 
finite size effect.

In addition, the nature of the counterion condensation is not 
well understood. By studying the condensation of counterions on a cylinder 
in three dimensions, and on a disc in two dimensions, it was shown that 
condensation is a second order phase transition \cite{Netz3,Netz2,burak}. 
However, to the best of our knowledge, the corresponding question has not been 
addressed for a three dimensional polyelectrolyte chain system.

In this paper, we simulate polyelectrolyte chains of different lengths and 
varying charge densities.  We argue that the sausage phase does not exist 
and is an artefact of studying very small chains. Our simulations also show
strong evidence that the 
counterion condensation is a second order transition accompanied by a 
divergence in the fluctuations of the number of non-bonded nearest 
neighbours of a monomer. In addition, we study aspects of the extended to 
collapsed transition.

\section{Model and simulation method}

We model the polyelectrolyte chain as spherical beads (monomers) 
connected through springs where each monomer carries a charge $qe$. 
Counterions are monovalent and are modelled as spherical beads, each carrying a charge 
$-qe$.  The polyelectrolyte chain and the counterions are assumed to be 
in a medium of uniform dielectric constant $\epsilon$. The potential 
energy due to the pair of particles $i$ and $j$ consists of three 
interactions.

Coulomb interaction: The electrostatic energy is given by 
\begin{equation}
U_{c}(r_{ij})=\frac{Zq^{2}e^{2}}{4\pi\epsilon r_{ij}},
\label{eq.1}
\end{equation}
where $Z=-1$ for monomer-counterion pairs and $Z=1$ otherwise, and 
$r_{ij}$ is the distance between particle $i$ and $j$.

Excluded volume interaction: The excluded volume interactions are 
modelled by the Lennard-Jones potential, which for two particles at a 
distance $r_{ij}$, is given by
\begin{equation}
U_{LJ}(r_{ij})= 4\epsilon_{ij} \left[ 
\left(\frac{\sigma}{r_{ij}}\right)^{12} 
-\left(\frac{\sigma}{r_{ij}}\right)^{6}\right],
\label{eq.2}
\end{equation}
where $\epsilon_{ij}$ is the minimum of the potential and $\sigma$ is 
the inter particle distance at which the potential becomes zero. We
use reduced units, in which the energy and length scales are specified
in units of $\epsilon_{ij}$ (counterion-counterion) and $\sigma$ respectively. 
The depth of the attractive potential 
$\epsilon_{ij}$ is chosen as 1.0 for  monomer-counterion and counterion-counterion pairs 
and $2.0$ for monomer pairs, while $\sigma$ is set to 1.0 for all pairs.  
We use the shifted Lennard-Jones potential in which 
$U_{LJ}(r_{ij})$ is set to zero beyond a cut off distance $r_{c}$. The 
value of $r_{c}$ equals 1.0 for monomer-counterion and 
counterion-counterion pairs and $2.5$ for monomer-monomer pairs. 
With this choice of the parameters, the excluded volume interaction is 
purely repulsive for all the pairs other than the monomer-monomer pairs. 
The effective short range attraction among the monomers mimics poor 
solvent conditions.  Other ways of realising poor solvent conditions may 
be found in Ref.~\cite{yethi}.

Bond stretching interaction: The bond stretching energy for pairs 
in the polymer that are connected directly through springs is given by
\begin{equation}
U_{bond}(r_{ij})=\frac{1}{2}k(r_{ij}-b)^{2},
\label{eq.3}
\end{equation}
where $k$ is the spring constant and $b$ is the equilibrium bond length. 
The values of $k$ and $b$ are taken as 500 and 1.12 respectively. This 
value of $b$ is close to the minimum of Lennard-Jones potential.

The relative strength of the electrostatic interaction is parametrised 
by a dimensionless quantity $A$:
\begin{equation}
A=\frac{q^{2}\ell_{B}}{b},
\label{eq.4}
\end{equation}
where $\ell_{B}$ is the Bjerrum length~\cite{Russel}, the length scale 
below which electrostatic interaction dominates thermal fluctuations.
\begin{equation}
\ell_{B}=\frac{e^{2}}{4\pi\epsilon k_{B}T},
\label{eq.5}
\end{equation}
where $k_{B}$ is the Boltzmann constant and $T$ is temperature. The 
thermal fluctuations dominate over the electrostatic interactions for 
very low values of $A$ and the counterions will be uniformly distributed 
in the solution. When $A$ is of order one, the electrostatic interaction
energy is comparable to the thermal energy and the counterions begin to
undergo Manning condensation\cite{man}. In our simulation, we vary $A$
from $0.055$ to $14.29$.

The equations of motion are integrated in time using the 
molecular dynamics simulation package LAMMPS~\cite{lammps1,lammps}. 
The simulations are carried out at constant temperature (T=1.0), maintained 
through a Nos\'{e}-Hoover thermostat 
(coupling constant $=0.1$)~\cite{Nose,Hoover}. The system is placed in a
cubic box with periodic boundary conditions. At the start of the
simulations, the configuration of the chain is randomly chosen and
the monovalent counterions are distributed uniformly throughout 
the  volume such that the charge neutrality is achieved. In our simulations,
the length $N$ of the chain is varied from 50 to 400, keeping the
overall particle density of
the system fixed at a constant value, $7.23~\times~10^{-6}$ particles per
$\sigma^3$. At this
density, there is no direct contact between the polyelectrolyte chain 
and any of its periodic images. 
We use the particle-particle/particle-mesh (PPPM)
technique~\cite{Hockney} to evaluate the energy and forces due to 
the long range Coulomb interactions. The time step of the integration
is chosen as $0.001$. Our simulation runs are divided into an equilibration run
($5\times10^6$ steps), followed by a production run ($6\times10^6$ steps). 
All the data shown in the plots are measurements over only the production
run ($1.2 \times 10^{5}$ configurations). The errors in the plots are
estimated by dividing the production into five statistically independent
blocks and measuring the standard deviation of the mean values of the
observable in each block \cite{gould}.

\section{Results and Discussion}

\subsection{Configurations of the polyelectrolyte chain}

\begin{figure}
\includegraphics[height=10cm]{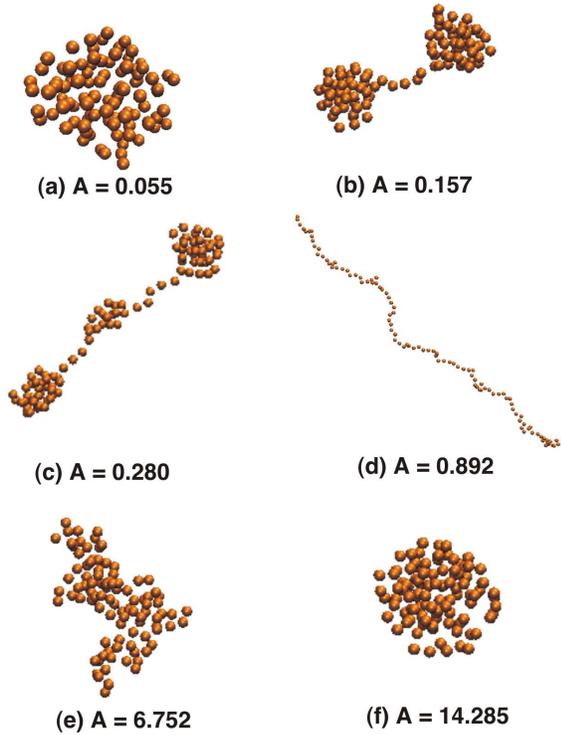}
\caption{Snapshots of the polyelectrolyte chain configurations for 
different values of $A$. For the sake of clarity, counterions are not 
shown and figures are not in the same scale. (a) Globular phase, 
(b) Dumbbell phase, (c) Pearl-necklace phase, (d) Extended phase, (e) 
Sausage phase and (f) Globular phase. }
\label{fig1}
\end{figure}

Snap shots of the
polyelectrolyte configuration for different values of $A$ are shown in 
Fig.~\ref{fig1}. For very small values of $A$, the electrostatic 
interactions can be ignored, and the polymer exists in a collapsed 
phase [Fig.~\ref{fig1}(a)]. As $A$ is increased, the globule breaks up into
two [Fig.~\ref{fig1}(b)] or more smaller globules [Fig.~\ref{fig1}(c)].
This pearl-necklace configuration becomes extended on further increasing $A$
[Fig.~\ref{fig1}(d)]. Counterion condensation is initiated and for
sufficiently large $A$, the polymer undergoes a collapse transition to form a
sausage phase \cite{jaya} [Fig.~\ref{fig1}(e)] or a spherical globule
[Fig.~\ref{fig1}(f)].

\subsection{Distribution of counterions}

\begin{figure}
\includegraphics[width=\columnwidth]{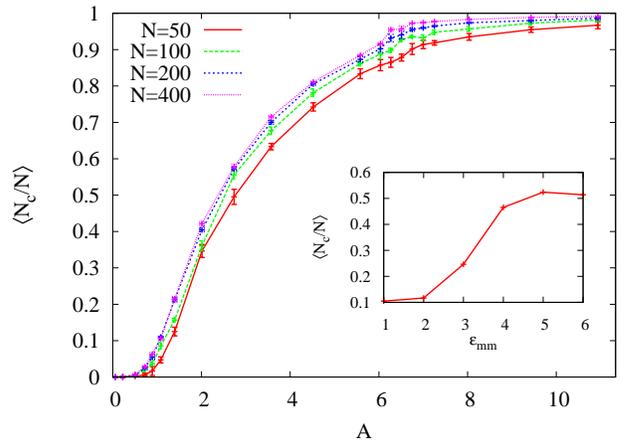}
\caption{Mean fraction of counterions $\langle N_c/N \rangle$ 
within a distance $2\ell_{B}$ from
the polyelectrolyte chain as a function of $A$. Inset: $\langle N_c/N
\rangle$ 
as a function of the monomer-monomer attraction energy
$\epsilon_{mm}$ for $N=100$ and $A=0.89$.}
\label{fig2}
\end{figure}

We consider a counterion to be condensed if its distance from any  
monomer is less than $2\ell_{B}$~\cite{wink}. Let $N_c$ be the number of condensed 
counterions and $N$ the number of monomers. The mean fraction of 
condensed counterions $\langle N_c/N \rangle$ as a function of $A$ is 
shown in Fig.~\ref{fig2} for different $N$. With the above definition of a
condensed couterion, it is observed that the counterion condensation 
starts slightly below $A=1$, consistent with the findings in 
Ref.~\cite{muthu3}. Note that for the case of a infinitely long and uniformly
charged cylinder, the classic Manning condensation occurs at $A=1$.

\begin{figure}
\includegraphics[width=\columnwidth]{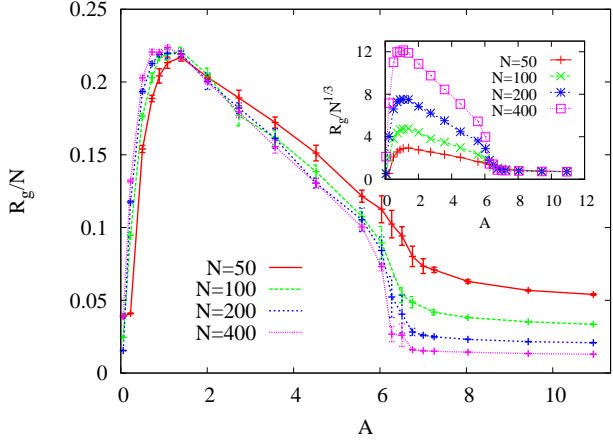}
\caption{Ratio of the radius of gyration $R_g$ to the number of monomers $N$
as a function of $A$. Inset: The data for $R_g/N^{1/3}$ collapse for
different $N$ for $A\gtrsim 6.25$.
}
\label{fig3}
\end{figure}

For a given value of $A$, the fraction of condensed counterions increases 
with increasing $N$ and reaches a limiting value. A similar result was
obtained for the good solvent case as well \cite{wink}.
The dependence of the fraction of condensed counterions on $N$ is closely
linked to the effective size or morphology of the polyelectrolyte chain.
We show that the longer chains have smaller relative size for
values of $A$ at which the counterion condensation occurs.  
The dependence of the relative size of the chain on $N$ can be studied
by calculating the radius of gyration $R_g$, defined as 
\begin{equation} 
R_{g}=\sqrt{\frac{1}{N}\sum_{i=1}^{N} (\vec{r}_i-\vec{r}_{cm})^{2}}, 
\label{eq.6} 
\end{equation} 
where $\vec{r}_i$ is the position of the $i^{th}$ particle and $\vec{r}_{cm}$ 
is the centre of mass of the chain. For fixed $A$, $R_g/N$, a measure
of the relative extension, decreases with $N$ in the region $A\gtrsim 0.89$ 
where the condensation occurs (see Fig.~\ref{fig3}). 

These two observations, the 
effective size being smaller for larger chains, and longer chains having a larger
fraction of condensed counterions (see Fig.~\ref{fig2}), are related.
We argue that a polymer with a smaller effective size has more condensed
counterions.
In the inset of Fig.~\ref{fig2}, we show that increasing $\epsilon_{mm}$
(depth of the Lennard-Jones potential of monomer-monomer pairs), keeping
other parameters fixed, results in increased condensation.
It is clear that increasing $\epsilon_{mm}$ can only 
result in the effective size of the polyelectrolyte chain becoming smaller. The
increased condensation may be due to counterions experiencing lower
electrostatic potential for a more compact chain.

We also observe that $R_g/N$ shows a jump around $A_c\approx6.25$. The jump is more
pronounced for longer chains. For $A>6.25$, radius of gyration
$R_{g}$ scales as $N^{1/3}$, indicating that the chain is in a
collapsed phase. We also observe a small kink in the fraction of
condensed ions (see Fig.~\ref{fig3}) around the same value of $A$ 
where $R_g/N$ shows a jump. 

\subsection{Number of non-bonded neighbours and the condensation transition}

\begin{figure}
\includegraphics[width=\columnwidth]{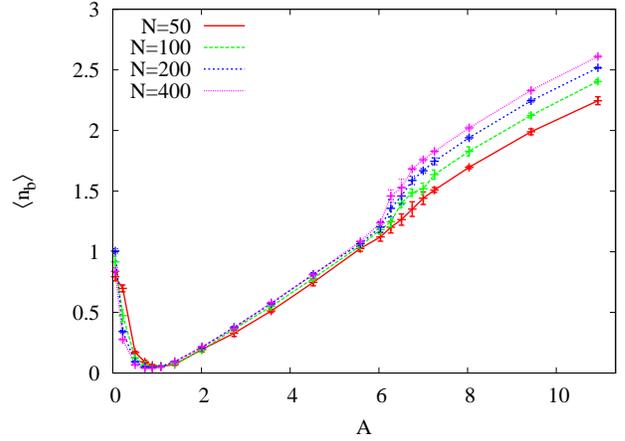}
\caption{The number of non bonded 
neighbours per monomer $\langle n_b\rangle$ as a function of $A$ for
different $N$}
\label{fig4}
\end{figure}

The number of non-bonded nearest neighbours of a monomer is a useful 
order parameter in studying the collapse transition of a neutral 
polymer~\cite{vander}. We study its behaviour for the polyelectrolyte 
chain. For a given monomer, a non-bonded neighbour is defined as any
monomer/counterion that is not connected to it by a bond and within a
distance $b$.  The variation of the mean 
number of non bonded neighbours per monomer $\langle n_b\rangle$ with 
$A$ is shown in Fig.~\ref{fig4}. $\langle n_b\rangle$ decreases when 
the polyelectrolyte chain goes from the initial collapsed phase to the 
extended phase. It has the minimum around $A'\approx 0.89$ where the chain is 
most extended. Close to $A_c\approx 6.25$, $\langle n_b\rangle$ has a jump 
with the jump size increasing with number of monomers $N$. This value of 
$ A $  corresponds to the collapse transition of the condensed 
polyelectrolyte (see inset of Fig.~\ref{fig3}).

\begin{figure} 
\includegraphics[width=\columnwidth]{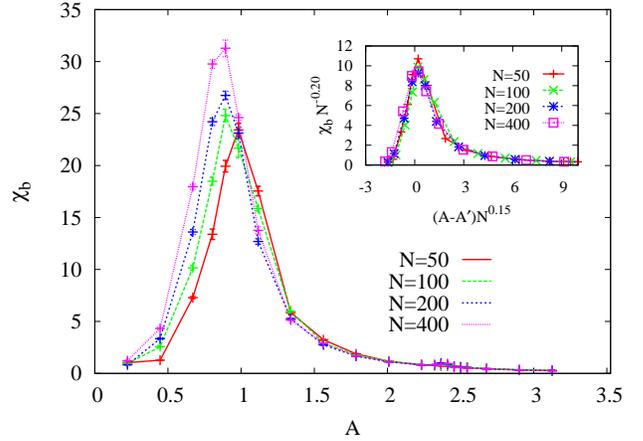} 
\caption{The relative 
fluctuation $\chi_b$ of the number of non bonded neighbours, as defined in 
Eq.~(\ref{eq.7}), as a function of $A$ for different $N$. Inset: Data
collapse when $\chi_b$ and $A$ are scaled as in
Eq.~(\ref{eq:chibscaling}), where $A'=1.08$ for $N=50$ and $A'=0.89$
otherwise.}
\label{fig5} 
\end{figure}

We also study the relative fluctuations $\chi_b$ of the number of 
non-bonded neighbours, where
\begin{equation}
\chi_{b}=\frac{N\left[ \langle n_{b}^{2}\rangle - \langle n_{b}
\rangle^{2} \right]}{\langle n_{b} \rangle^{2}}. 
\label{eq.7}
\end{equation}
It has a peak  around $A'\approx0.89$ (see Fig.~\ref{fig5}).
The increasing peak height with the number of monomers
$N$, indicates a divergence in the thermodynamic limit. 
This critical value $A'$ 
corresponds to the onset of condensation of the counterions. At this value of
$A$, the polyelectrolyte chain is fully extended and hence this condensation
is reminiscent of Manning condensation. The data for
different values of $N$ can be collapsed using the scaling form
\begin{equation}
\chi_b \approx  N^{\phi_1} f\left[(A-A') N^{\phi_2}\right], ~N \gg 1,
\label{eq:chibscaling}
\end{equation}
where $\phi_1$ and $\phi_2$ are scaling exponents. Data collapse is seen for 
$\phi_1 \approx 0.20$ and $\phi_2\approx 0.15$ (see inset of 
Fig.~\ref{fig5}). Divergence of $\chi_b$ with data collapse is a strong 
indication of condensation being a continuous phase transition. Earlier
discussion of the order of the Manning like  condensation has been 
restricted only to model systems of cylinder in three dimensions and disc 
in two dimensions, and a similar continuous transition 
has been reported in such systems~\cite{Netz2,Netz3,burak}.

\subsection{\label{sec:collapse} Transition from extended phase to 
collapsed phase}

To quantify the transition from the extended phase to the collapsed 
phase, the electrostatic energy per monomer $E_c$ and its fluctuations 
are calculated. Only monomer-monomer pairs are used for calculating 
these quantities since we are interested in the configuration of 
the polyelectrolyte chain. The relative fluctuation $\chi_c$ in the 
electrostatic energy is defined as
\begin{equation}
\chi_c=\frac{N\left[\langle E_c^{2}\rangle-\langle E_c
\rangle^{2}\right]}{\langle E_c \rangle^{2}}.
\label{eq.10}
\end{equation}
 
\begin{figure}
\includegraphics[width=\columnwidth]{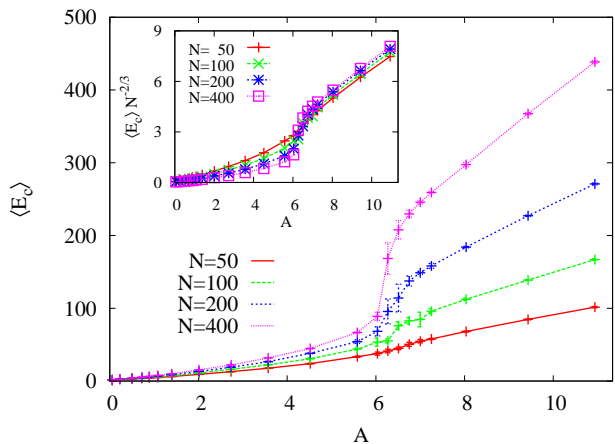}
\caption{
Mean Coulomb energy per monomer $\langle E_c \rangle$
as a function of $A$ for different $N$. Inset: $\langle E_c \rangle \sim
N^{2/3}$ in the collapsed phase.}
\label{fig6}
\end{figure}

The mean electrostatic energy per monomer
$\langle E_c \rangle$ increases with $A$ and has an abrupt jump at
$A_c\approx 6.25$ (see Fig.~\ref{fig6}). 
The jump is more pronounced for higher values of 
$N$. In the collapsed phase, the shape of the chain is roughly spherical and
hence $\langle E_c \rangle$ should scale as $N^{2/3}$ \cite{landaubook}. 
For $A> A_c$, we confirm that $\langle E_c \rangle$ scales 
as $N^{2/3}$ (see inset of Fig.~\ref{fig6}). The  scaling of $\langle E_c
\rangle$ as $N^{2/3}$ along with the scaling of $R_g$ as $N^{1/3}$ (see inset
of Fig.~\ref{fig3}) in this regime confirm that the polymer
is indeed in a collapsed configuration.

\begin{figure}
\includegraphics[width=\columnwidth]{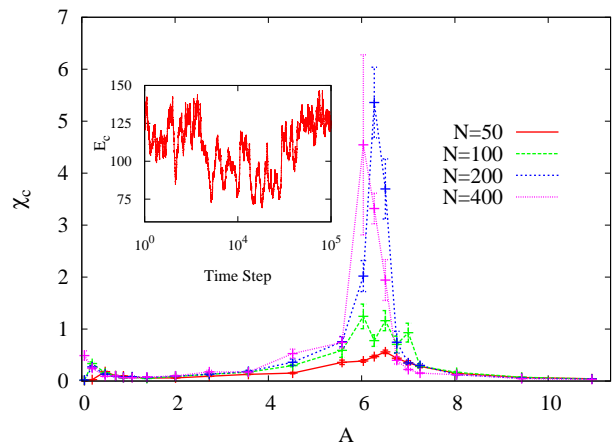}
\caption{Fluctuation in the Coulomb energy per monomer $\chi_c$
as a function of $A$. Inset: Time series of the electrostatic energy
per monomer, for $N=200$, and $A=6.50$.}
\label{fig7} 
\end{figure}

The variation of the relative fluctuation $\chi_c$ with $A$
is shown in Fig.~\ref{fig7}.
$\chi_c$ has  a peak at $A_c$, close to the value of $A$ at which
$\langle E_c \rangle$ has a discontinuity.
The peaks are not resolved to the
desired accuracy. This is because, close to the transition point,
the polyelectrolyte chain fluctuates between the extended
and the collapsed phases during the time evolution.
The inset of Fig.~\ref{fig7} 
shows a sample  time series of $E_c$, near the transition point, for $N=200$,
where $E_c$ fluctuates roughly between two values.
This behaviour of $\chi_c$ coupled with the sharp rise in $\langle E_c
\rangle$ (see Fig.~\ref{fig6})  suggests a first
order phase transition.
Our findings are consistent with the previous theoretical
results \cite{brilliantov} 
that the extended to collapse transition of a polyelectrolyte chain is first 
order.

Similar divergence in the fluctuation of the electrostatic energy 
per monomer has been
observed for the transition of a polyelectrolyte chain,
in a poor solvent, from the initial globular phase to the dumbbell and
trimbell phases in the absence of explicit counterions \cite{Lyulin}.

\subsection{Existence of Sausage phase}{\label{sec:asphericity}

The sausage phase was defined in Ref.~\cite{jaya} as a collapsed phase where
mean asphericity is non-zero. We show below  that
the transition associated with change in asphericity coincides with the
collapse transition.
We define asphericity $Y$ as
\begin{equation}
Y=\left\langle\frac{\lambda_{1} - \frac{\lambda_{2} + \lambda_{3}}{2}} 
{\lambda_{1}+\lambda_{2} + \lambda_{3}}\right\rangle,
\label{eq.8}
\end{equation}
where $\lambda_{1,2,3}$ are the eigenvalues 
of the moment of inertia tensor with $\lambda_{1}$ being the largest 
eigenvalue. The moment of inertia tensor $G$ is 
\begin{equation}
G_{\alpha\beta}=\frac{1}{N}\sum_{i=1}^{N}r_{i\alpha}r_{i\beta},
\label{eq.9}
\end{equation}
where $r_{i\alpha}$ is the $\alpha^{th}$ component of the 
position vector $\vec{r}_{i}$. Asphericity $Y$ is zero for a
sphere (collapsed globule) and one for a linear rod (extended configuration).
For all other configurations, it has a value between zero and one.

The variation of asphericity with $A$ for different $N$ is shown
in Fig.~\ref{fig8}. For very small values of $A$, asphericity increases
corresponding to the extension of the initial collapsed phase. In the
extended region $0.89 \lesssim A\lesssim 6.25$, 
asphericity increases with $N$ and tends to
one for large $N$.  For $A \gtrsim 6.25$, asphericity decreases to zero
with $N$ as a power law (see inset of
Fig.~\ref{fig8}). Thus, for large $N$, asphericity jumps from
one to zero as $A$ crosses $A_c \approx 6.25$. The value of the
transition point coincides with the extended to collapsed transition discussed in Sec.~\ref{sec:collapse}. As
there is no second transition in asphericity or in $\langle E_c \rangle$, we conclude that
the sausage phase suggested in Ref.~\cite{jaya} is identical with the
collapsed phase and is not a different phase.
\begin{figure}
\includegraphics[width=\columnwidth]{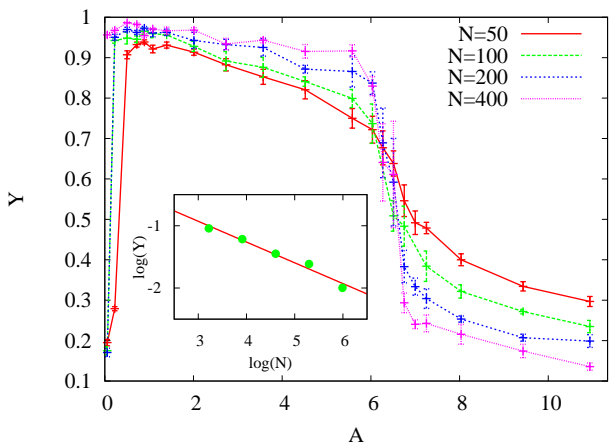}
\caption{Asphericity $Y$ (as defined in Eq.~(\ref{eq.8})  as a function of 
$A$. 
Inset: Variation of asphericity $Y$ with chain length $N$ for $A=10.93$. 
The slope of the straight line is $-0.33$.}
\label{fig8}
\end{figure}

\section{Conclusions}

Using molecular dynamics simulations, we studied 
the phase diagram and phase transitions between the
phases of a single polyelectrolyte chain in a poor solvent. The counterions
were taken care of explicitly while the solvent was implicit. The dependence
of various physical quantities on $A$, a dimensionless number that
parametrised the strength of the electrostatic interaction, were calculated
for different chain lengths $N$. Our main results are summarised below.

We quantified the transition associated with the condensation of
counterions on the extended chain. The fluctuations of the non-bonded
neighbours $\chi_b$ diverges at $A'\approx 0.89$ with
exponents that indicate a continuous transition. While the value of the
exponents we obtained are not accurate, they indicate that data for
different system sizes can be explained by one scaling function. It would be
interesting to study the number of non-bonded neighbours and its
fluctuations for the good solvent problem as well as in the presence of
salt.

We also analysed the sausage phase introduced in Ref.~\cite{jaya}
as a new intermediate phase
between the extended  and collapsed phases.
We show that the transition associated with the discontinuity in
asphericity and that associated with the extended to collapsed
transition occur, within
numerical error, at the same value of $A$ ($A_c \approx 6.25$). 
This, in conjunction with the fact
that there is no second transition for either quantity, strongly
suggests that the sausage phase does not exist independent of the collapsed phase,
and was probably an artefact of earlier simulations \cite{jaya} of
chains of small size.

The extended to collapsed transition of the condensed polymer chain was
also quantified. The abrupt jump in $R_g/N^{1/3}$,  in the mean 
electrostatic energy $\langle E_c\rangle$, and mean number of non-bonded 
nearest neighbours $\langle n_b \rangle$, transition of the chain
between extended phase and collapsed phase near the transition point, 
combined with  
the peaks in the fluctuations of the
electrostatic energy $\chi_c$, are evidences
for a first order transition. This is consistent with the theoretical
prediction in Ref.~\cite{brilliantov}.

In addition, we find that
the fraction of condensed counterions increases with $N$ for fixed $A$ in the
extended phase. By studying its dependence on $\epsilon_{mm}$ and the dependence of $R_g/N$ on
$N$, we argued that the increase in the fraction  is related to longer
chains having a smaller relative extension.

\end{document}